\documentclass[a4paper,11pt]{article}
\usepackage{jcappub} % for details on the use of the package, please see the JINST-author-manual
\usepackage{lineno}
\usepackage{subcaption}
\arxivnumber{2406.12281} % Only if you have one
\title{\boldmath Projected gravitational wave constraints on primordial black hole abundance for extended mass distributions}

\author[1]{G. L. A. Dizon\note{Corresponding author.}}
\author{and R. C. Reyes}
\affiliation{National Institute of Physics,\\University of the Philippines - Diliman,\\
Quezon City, Philippines}

\emailAdd{gldizon@nip.upd.edu.ph}

\abstract{We investigate the projected minimum constraints set by next-generation gravitational wave detectors Einstein Telescope and LISA on the abundance of primordial black holes relative to dark matter from both resolvable mergers and the stochastic gravitational wave background (SGWB) for extended primordial black hole mass distributions. We consider broad power law distributions for a range of negative and positive exponents $\gamma$ and top-hat distributions (with $\gamma=0$) and use the IMRPhenomXAS waveforms to simulate binary sources up to mass ratios $q_\mathrm{max} = 1000$ and redshifts $z=300$. Our results suggest that accounting for extended mass distributions have the most apparent impact when considering mergers at high redshifts $z > 30$, for which the constraint curves have broader mass windows and shift to higher abundances compared to when a monochromatic distribution is assumed; on the other hand, constraints from low-redshift mergers and the SGWB do not change much with the assumed mass distribution. At high redshifts, astrophysical black holes are not expected to contribute significantly, providing possible smoking-gun evidence for PBHs. Constraints derived from LISA and ET observations would complement each other by probing different PBH mass windows and this holds for the extended mass distributions studied.}

\begin{document}
\maketitle
\flushbottom

\section{Introduction}
\label{sec:intro}
Primordial black holes (PBHs) have long been considered a possible solution to the dark matter problem, with interest resurging due to their possible contribution to merger signals in gravitational wave (GW) detectors such as LIGO~\cite{PhysRevX.11.021053}, along with more recent stochastic GW background (SGWB) measurements from pulsar timing arrays like NANOGrav probing potential PBH formation mechanisms~\cite{universe9040157, Afzal_2023}. Constraints on PBH population abundance (relative to dark matter) $f_\mathrm{PBH}$ have been derived from various observations~\cite{Carr_2021, 10.3389/fspas.2021.681084} such as microlensing~\cite{PhysRevD.99.083503, PhysRevD.101.063005, PhysRevLett.111.181302, Griest_2014, Alcock_2001, PhysRevLett.121.141101, PhysRevD.86.043001, refId0, PhysRevD.94.063530}, CMB temperature data~\cite{PhysRevResearch.2.023204}, as well as from dynamical~\cite{Carr_1999, 10.1093/mnras/sty1204, PhysRevD.94.063530} and structural~\cite{PhysRevLett.123.071102} considerations for observed cosmological structures. While these constraints are well-established for both monochromatic \cite{10.3389/fspas.2021.681084} and extended mass distributions \cite{Bellomo_2018,PhysRevD.96.023514,PhysRevD.94.063530, Kuhnel:2017pwq}, there is still a sizeable lack of constraints derived from GW data in the literature. Current GW-based constraints from existing~\cite{PhysRevD.103.023026, Raidal_2017, Hutsi_2021, PhysRevD.110.023040} and projected observations~\cite{De_Luca_2021, Ng_2022} either largely assume a monochromatic (or narrow-width) initial mass distribution of PBHs upon formation, or even with consideration of broad distributions only account for symmetric-mass binary mergers~\cite{SPP-2022-1B-03}. 

This paper extends part of existing work by De Luca, et al.~\cite{De_Luca_2021} in projecting minimum PBH abundance constraints using simulated LISA and ET detector models. While they  only report results for monochromatic mass distributions, here we consider extended mass functions with a simple power law form $\psi(M) \propto M^{-\gamma}$ for both positive and negative values of $\gamma$, as well as for $\gamma=0$, corresponding to a top-hat distribution; we allow for values of $\gamma$ to admit both very narrow and very broad mass function widths. In order to fully capture the effect of such a range of mass functions, we consider binary mergers with mass ratios up to $q_\mathrm{max} = 1000$, the maximum allowed by the range of validity of our assumed waveform approximant. We derive constraints from both resolvable mergers and the SGWB. For the former, we split our analysis between low-redshift $(z \leq 30)$ and high-redshift $(z > 30)$ mergers to investigate the impact of extended mass functions on the detectibility of mergers at different redshift regimes. 

%\textcolor{blue}{Our results show that the effect of a broad mass distribution is most apparent when considering high-redshift mergers, although }

We structure this paper as follows. Section~\ref{sec:modeling} discusses the setup and assumptions on both PBH sources and detector models used to forecast constraints as well as the mass distributions considered. Section~\ref{sec:observables} outlines how the constraining observables are calculated, namely the expected detectable merger event rate for resolvable mergers and the expected SGWB spectrum. We present our constraint procedure as well as our derived abundance constraints in Section~\ref{sec:measurable}; additionally, we report systematic tests on the effect of varying $q_{\rm max}$. Lastly, we discuss our conclusions and outlook in Section~\ref{sec:conclu}.

\section{Source and detector models}
\label{sec:modeling}

\subsection{Parameter initialization and GW waveforms}
\label{subsec:waveforms}
\begin{figure}
    \centering
    \includegraphics{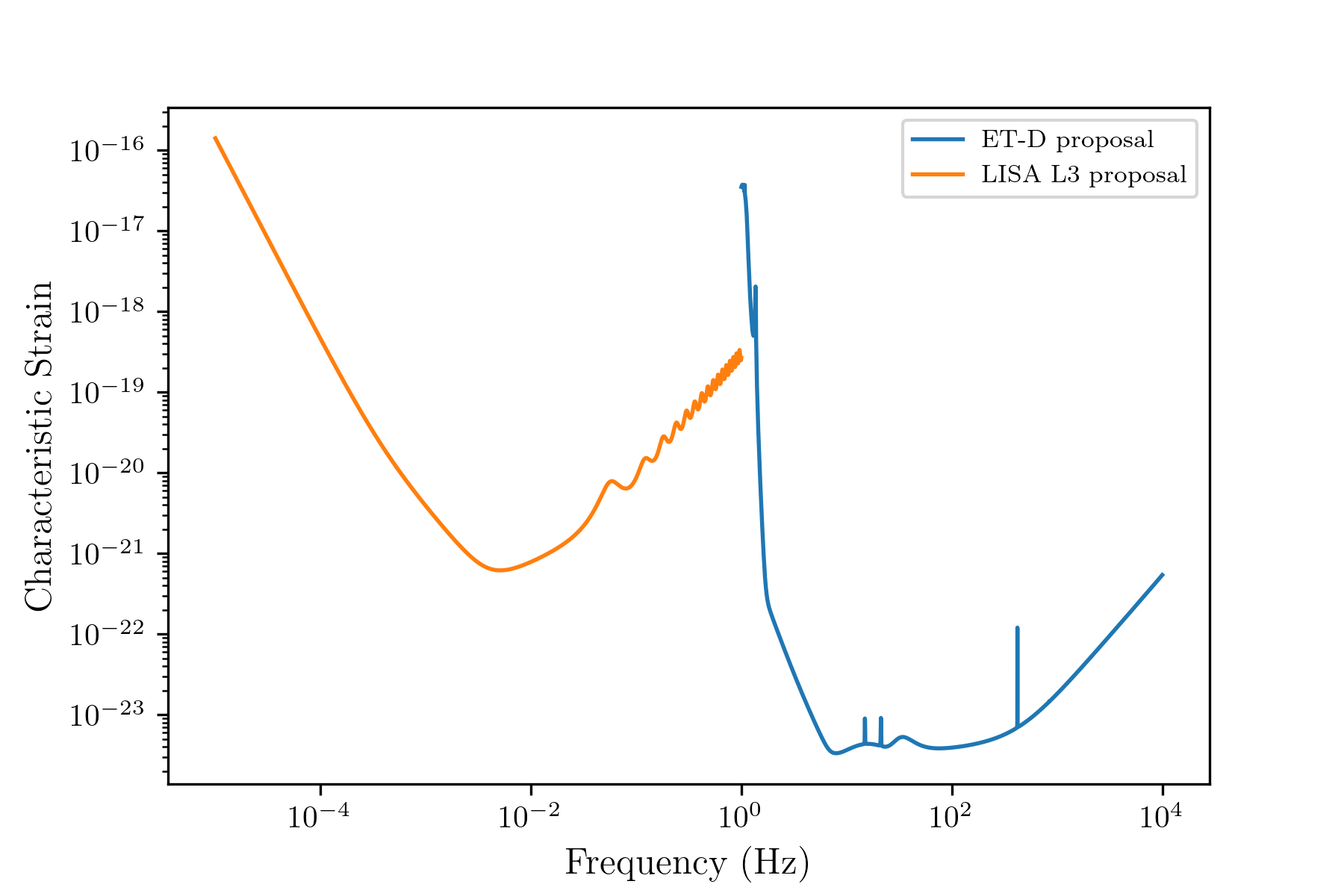}
    \caption{Proposed sensitivity curves for the Einstein Telescope~\cite{Hild_2011} and LISA~\cite{amaroseoane2017laser}, based off existing design proposals.}
    \label{fig:detectors}
\end{figure}

All parameter initialization and source-detector simulations are facilitated with the use of Python package {\ttfamily{gwent}}~\cite{Kaiser:2021}. For detectors, we chose to simulate the LISA L3 proposal~\cite{amaroseoane2017laser} and the Einstein Telescope (ET) design proposal D~\cite{Hild_2011}. This choice of detectors allow us to cover at least four decades of the GW frequency band. Figure~\ref{fig:detectors} shows the sensitivity curves for both LISA and ET (orange and blue curves, respectively). 

%\textcolor{blue}{For PBH binary sources, we cover the full scope allowed by our choice of detectors, with the lowest total binary mass limit at $M_\mathrm{tot} = M_1 + M_2 \sim 10^{-6}\,M_\odot$,} where $M_1$ and $M_2$ are the component masses of the binary. 

For resolvable PBH binary sources, we cover the full range of total binary masses $M_\mathrm{tot} = M_1 + M_2$ detectable by LISA and ET, respectively, namely: $M_\mathrm{tot} = 1$--$10^9\,M_\odot$ for LISA and $M_\mathrm{tot} = 10^{-6}$--$10^6\,M_\odot$ for ET. Contributions from binaries with $M_\mathrm{tot}$ outside these ranges are treated as negligible, as they have insignificant contribution to the SNR (as shown in Fig.~\ref{fig:SNR_q} below). To include unresolved sources that fall within the SGWB window, we extend LISA's range to $M_\mathrm{tot} = 10^{-6}$--$10^9\,M_\odot$ in the case of the SGWB constraints. We also assume that the sources have no spin, i.e., set the component spin parameters $\chi_1$ and $\chi_2$ to zero, in line with the assumption that all PBHs form from almost perfectly spherically symmetric density peaks\footnote{This assumption does not necessarily hold for PBHs formed during a matter-dominated epoch. See Ref.~\cite{PhysRevD.96.083517} for further reading.}~\cite{10.1093/ptep/ptx087}. 

The GW waveform package typically employed in PBH studies is IMRPhenomD~\cite{PhysRevD.93.044006, PhysRevD.93.044007}, which models black hole binaries with non-precessing spins up to a maximum mass ratio of $q_\mathrm{max} = 18$. The reasons for its ubiquity are twofold: most present PBH merger studies only deal with monochromatic sources with symmetric binaries $(q=1)$, and the package has a Python implementation pyIMRPhenomD~\cite{PhysRevD.108.023022} which significantly optimizes calculation for Python codes. Given our interest in modeling extended mass distributions, we instead use IMRPhenomXAS as our waveform model. IMRPhenomXAS is also used for non-precessing black hole binaries, but it is also stated as an improvement over IMRPhenomD, covering mass ratios up to $q_\mathrm{max} =1000$~\cite{PhysRevD.102.064001}. 
%For brevity, we will henceforth refer to IMRPhenomD and IMRPHenomXAS as D waveforms and XAS waveforms, respectively.

In contrast to existing actual and projected constraints to PBH abundance from GW observations~\cite{Luca_2020, De_Luca_2021}, which assume narrow mass functions, essentially including only binaries with relatively low mass ratios ($q \lesssim 10$), we consider power law mass functions with mass ranges allowed by the scope of validity of the IMRPHenomXAS waveform models and include systems with a range of binary mass ratios $q = 1$--$1000$, allowing us to probe the intermediate mass ratio inspiral (IMRI) regime of PBH mergers. 
%In order to properly model black hole binaries with these mass ratios, we require the use of appropriate waveforms to model the GW signals from these sources, as well as mass functions that accommodate these ranges. We discuss the latter in Section~\ref{subsec:mass_functions}.

\subsection{Calculating detector SNR}
\label{subsec:dets}

\begin{figure}
    \centering
    \includegraphics[width=0.8\linewidth]{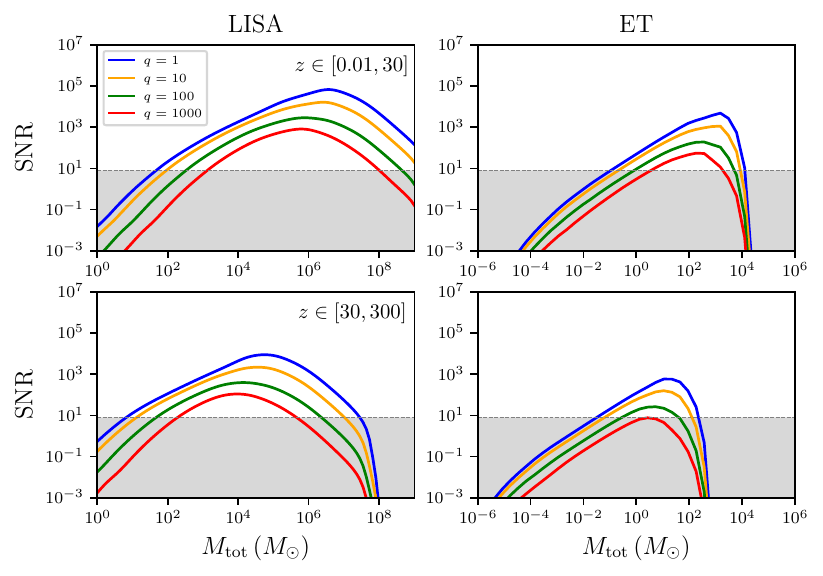}
    \caption{Detector SNRs as a function of total mass $M_\mathrm{tot}$ for fixed mass ratios $q = 1,\,10,\,100,\,1000$, integrated over low (top panels) and high (bottom panels) redshift ranges, for LISA (left panels) and ET (right panels). The shaded region represents $\mathrm{SNR} \leq 8$, lower than the threshold for a significant detection.}
    \label{fig:SNR_q}
    
    \includegraphics[width=0.8\textwidth]{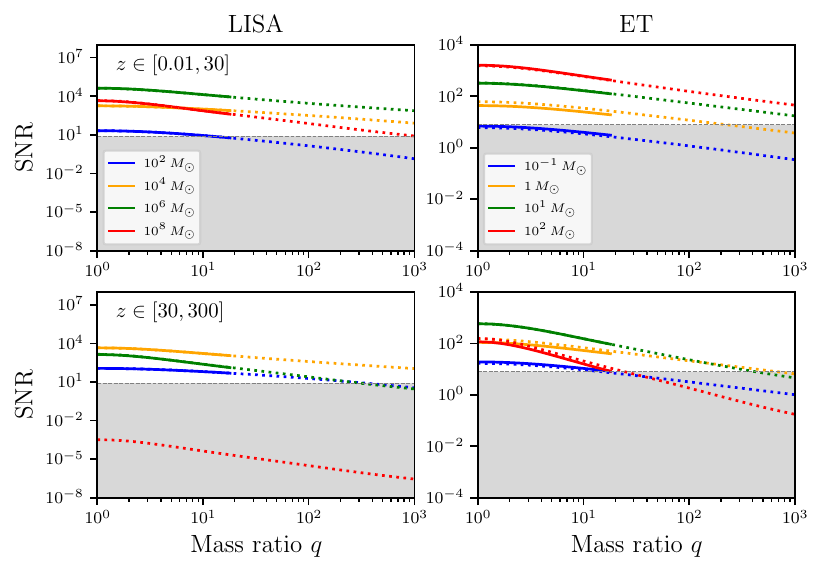}
    \caption{Detector SNRs as a function of binary mass ratio $q$ for fixed total masses $M_\mathrm{tot} = 10^2, 10^4, 10^6$ and $10^8\, M_\odot$ for LISA (left panels) and $M_\mathrm{tot} = 10^{-1}, 1, 10,$ and $100\,M_\odot$ for ET (right panels), integrated over low (top panels) and high (bottom panels) redshift ranges. Solid curves show SNR curves for IMRPhenomD waveforms (up to $q_\mathrm{max} = 18$), while dotted curves show SNR for IMRPhenomXAS waveforms (up to $q_\mathrm{max} = 10^3$). The shaded region represents $\mathrm{SNR} \leq 8$, lower than the threshold for a significant detection.}
    \label{fig:SNR}
\end{figure}

Detector SNR $\rho_\mathrm{det}$ is a function of source parameters (binary mass components, spin components, redshift), source waveform models (e.g., IMRPhenomD, IMRPhenomXAS), and detector sensitivity curves. The SNR calculation is also facilitated through the use of {\ttfamily{gwent}}. Source and detector parameters are as described above in Section~\ref{subsec:waveforms}. We calculate the SNR in a $M_\mathrm{tot}-q$ grid, integrating over a range of redshifts, from $z=0.01$--$30$ for the low-redshift range and from $z=30$--$300$ for the high-redshift range.

Figure~\ref{fig:SNR_q} shows SNR curves for both LISA and ET detectors as a function of total mass $M_{\rm tot}$, for fixed mass ratios $q = 1, 10, 100,$ and $10^3$, for the low-redshift and high-redshift ranges (top and bottom panels, respectively). The shaded regions represent $\mathrm{SNR} \leq 8$, lower than the threshold for a significant detection. These plots show the varying ranges in $M_{\rm tot}$ that LISA and ET can detect and justifies our choice of included mass ranges for the two detectors, as specified above in Sec.~\ref{subsec:waveforms}.

Additionally, Figure~\ref{fig:SNR} shows SNR as a function of binary mass ratio $q$ (dotted curves) for fixed total masses $M_\mathrm{tot} = 10^2, 10^4, 10^6$ and $10^8\, M_\odot$ for LISA (left panels) and $M_\mathrm{tot} = 10^{-1}, 1, 10,$ and $100\,M_\odot$ for ET (right panels), for the low-redshift and high-redshift ranges (top and bottom panels, respectively). Here, we have selected masses for illustration based on the respective mass windows of the detectors. The trend in SNR with varying total mass is consistent with the inverted U-shape of the mass window curves in Fig.~\ref{fig:SNR_q}, as expected. As another check, we also show the SNR curves calculated using the IMRPhenomD waveform, over the scope of its validity ($q \le 18$) (solid curves). As expected, the curves closely match the extended curves (using the IMRPhenomXAS waveform) over the regime where they overlap-- with the exception of the SNR curve for $M_{\rm tot}=10^8\, M_\odot$ at high redshift, where the SNR using the IMRPhenomD waveform is much lower (not shown in the plot); both curves fall well below the SNR threshold.

% For LISA: show M_tot = 10^2, 10^4, 10^6, 10^8
% For ET: show M_tot = 10^-1, 1, 10, 100

%As expected, D and XAS waveforms are generally in agreement up to mass ratios $q=18$. This is much more clear in the LISA SNRs, as there is near-unnoticeable deviation between the SNR of the waveforms for the case of ET at low redshifts and higher masses, as seen in the top left panel of Figure~\ref{fig:SNR}. \textcolor{blue}{The SNR curves for XAS waveforms extend smoothly up to $q = 1000$, monotonically decreasing for increasing mass ratios. Figure~\ref{fig:SNR_q} shows much more clearly how the XAS SNR varies with $q$ across each detector's mass range at different redshift intervals. Generally, individual binary mergers are still resolvable above the SNR = 8 threshold even at intermediate and extreme mass ratios, although the mass interval where this works narrows down with increasing $q$ and redshift.} 

\subsection{Mass distributions}
\label{subsec:mass_functions}
In this work, we consider bounded power law PBH mass distributions as well as monochromatic distributions for baseline comparison. While there are various choices for the normalization of the mass function, we choose to set $\int\psi(M)\,\mathrm{d}M = 1$, such that $\psi(M)$ has units of $[M_\odot^{-1}$]. We also define mass-averaged quantities $\langle X(M) \rangle$, following Ref.~\cite{raidal2024formationprimordialblackhole}, as
\begin{equation}
\label{eq:expectation_value}
    \langle X(M) \rangle = \frac{\int\,\mathrm{d}M\,\psi(M)XM^{-1}}{\int\,\mathrm{d}M\,\psi(M)M^{-1}}.
\end{equation}
In particular, the calculations for the reference mass $M_c \equiv \langle M \rangle$ and $\langle M^2 \rangle$ follow from this definition.

\subsubsection{Monochromatic mass distribution}
The monochromatic mass distribution is simply defined by the Dirac delta distribution
\begin{equation}
    \label{eqref:mono}
    \psi_\mathrm{mono}(M; M_c) = \delta_D(M - M_c),
\end{equation}
where $M_c$ is the central mass of the distribution. This distribution is considered primarily as a simplification, and is useful when considering symmetric mass binary GW sources. Trivially, the average mass and average squared mass of this distribution are just $\langle M \rangle = M_c$ and $\langle M^2 \rangle = M_c^2$, respectively.

\subsubsection{Bounded power law mass distribution}
\label{subsubsec:dist_plaw}

The bounded power law mass distribution is given by~\cite{Bellomo_2018}:
\begin{equation}
    \label{eqref:pl}
    \psi_{\mathrm{PL}}(M; M_\mathrm{min}, M_\mathrm{max}, \gamma) =
    \begin{cases}
        \mathcal{N}_\mathrm{PL}M^{\gamma - 1}, &\quad \mathrm{when}\, M \in [M_\mathrm{min}, M_\mathrm{max}], \\
        0, &\quad \mathrm{otherwise,}
    \end{cases}
\end{equation}
where $\mathcal{N}_\mathrm{PL}$ is the normalization factor given by
\begin{equation}
    \mathcal{N}_\mathrm{PL} = 
    \begin{cases}
        \frac{\gamma}{M_\mathrm{max}^\gamma - M_\mathrm{min}^\gamma}, &\quad \gamma \neq 0 \\
        \frac{1}{\log(M_\mathrm{max}/M_\mathrm{min})}, &\quad \gamma = 0.
    \end{cases}
\end{equation}
Here, $M_\mathrm{min}$ and $M_\mathrm{max}$ are the minimum and maximum mass cut-offs of the distribution, and are related to each other by $q_\mathrm{max} = M_\mathrm{max}/M_\mathrm{min}$, i.e., the maximum possible mass ratio for that distribution. As noted in Sec.~\ref{subsec:waveforms}, we set $q_{\rm max}=1000$ based on the scope of validity of our assumed waveform model. Later in Sec.~\ref{subsec:varying_qmax}, we investigate the effect of varying $q_\mathrm{max}$ on our derived PBH abundance constraints. The exponent $\gamma$ relates to the cosmological equation of state during the PBH formation epoch~\cite{1975ApJ...201....1C}. A power law mass distribution is also formulated by Ref.~\cite{DELUCA2020135550}, which sets the exponent at $\gamma = -1/2$. This particular form of the mass distribution arises from assuming that collapsing primordial fluctuations form a broad and flat power spectrum in $k$-space. 

From Eq.~\ref{eq:expectation_value}, we may compute for $\langle M \rangle$ and $\langle M^2 \rangle$ for this bounded power law distribution. We find that
\begin{equation}
\label{eq:M_c}
   M_c \equiv \langle M \rangle =
   \begin{cases}
       \frac{q_\mathrm{max} - 1}{\log q_\mathrm{max}}M_\mathrm{min}, &\quad \gamma = 1, \\
       \frac{q_\mathrm{max} \log q_\mathrm{max}}{q_\mathrm{max} - 1}M_\mathrm{min}, &\quad \gamma = 0,\\
       \frac{\gamma - 1}{\gamma}\frac{q_\mathrm{max}^\gamma - 1}{q_\mathrm{max}^{\gamma - 1} - 1}M_\mathrm{min}, &\quad \mathrm{otherwise,}
   \end{cases}
\end{equation}
where we have reduced the general expression for the special cases, $\gamma=0$ and 1. Similarly, we find that $\langle M^2 \rangle$ is given by
\begin{equation}
    \langle M^2 \rangle = 
    \begin{cases}
        \frac{q_\mathrm{max}^2 - 1}{2\log q_\mathrm{max}}M_\mathrm{min}^2, &\quad \gamma = 1, \\
       \frac{2q^2_\mathrm{max} \log q_\mathrm{max}}{q^2_\mathrm{max} - 1}M^2_\mathrm{min}, &\quad \gamma = -1,\\
       \frac{\gamma - 1}{\gamma + 1}\frac{q_\mathrm{max}^{\gamma+1} - 1}{q_\mathrm{max}^{\gamma - 1} - 1}M^2_\mathrm{min}, &\quad \mathrm{otherwise,}
    \end{cases}
\end{equation}
where we have reduced the general expression for the special cases, $\gamma=-1$ and 1.

Any mass distribution can be characterized by a dimensionless width parameter $\sigma_\mathrm{PBH}$ characterizing the spread of the distribution, given by $\sigma_\mathrm{PBH}^2 = (\langle M^2 \rangle / \langle M \rangle^2) - 1$. For the case of the bounded power law distribution defined in Eq.~\ref{eqref:pl}, the width $\sigma_\mathrm{PBH}$ simply depends on $|\gamma|$ and $q_{\rm max}$. Table~\ref{tab:sigma_gamma} lists values of $\sigma_\mathrm{PBH}$ for selected values of $|\gamma|$ and $q_{\rm max}$, for reference. We note that $\sigma_\mathrm{PBH}$ attains its largest value when $\gamma = 0$, given by
\begin{equation}
    \sigma_\mathrm{PBH}^2 = \frac{(q_\mathrm{max}-1)^2}{q_\mathrm{max}\log^2 q_\mathrm{max}} - 1.  
\end{equation}
For our chosen value of $q_\mathrm{max}=1000$, $\sigma_{\mathrm{PBH},\gamma=0} \approx 4.46$. Another special case is for $\gamma = \pm 1$, where the expression reduces to
\begin{equation}
    \sigma_\mathrm{PBH}^2 = \frac{\log q_\mathrm{max}}{2}\left(\frac{q_\mathrm{max} + 1}{q_\mathrm{max} - 1}\right) - 1.
\end{equation} 
For our chosen value of $q_{\rm max}=1000$, $\sigma_{\mathrm{PBH},\gamma=\pm 1} \approx 1.57$. For $|\gamma| < 1$, $\sigma_\mathrm{PBH}$ ranges from $\sigma_{\mathrm{PBH},\gamma=\pm 1}$ to $\sigma_{\mathrm{PBH},\gamma=0}$. On the other hand, for $|\gamma| > 1$, $\sigma_\mathrm{PBH} < \sigma_{\mathrm{PBH},\gamma=\pm 1}$ and approaches zero as $|\gamma| \rightarrow \infty$, reducing to the monochromatic case. 

\begin{table}[]
\centering
\begin{tabular}{|r|r|r|r|}
\hline
& \multicolumn{3}{c|}{$\sigma_\mathrm{PBH}$}\\\hline
\multicolumn{1}{|c|}{$|\gamma|$} & $q_\mathrm{max} = 10$ & $q_\mathrm{max} = 100$ & $q_\mathrm{max} = 1000$ \\
\hline
10                             & 0.1005                & 0.1005                 & 0.1005                  \\
1                              & 0.6381                & 1.1615                 & 1.5687                  \\
0.1                            & 0.7255                & 1.8911                 & 4.3926                  \\
0.01                           & 0.7265                & 1.9029                 & 4.4619                  \\
0                              & 0.7265                & 1.9030                 & 4.4626   \\\hline              
\end{tabular}
\caption{$\sigma_\mathrm{PBH}$ for selected values of $|\gamma|$ at different $q_\mathrm{max}$, up to four decimal places.}
\label{tab:sigma_gamma}
\end{table}

%\footnote{Contrast with Ref.~\cite{PhysRevD.96.023514}, which defines its extended mass distribution moments as $\langle \log^n M \rangle$.} 
%While we do not write out the explicit form of $\sigma$ for the general case, we point out two special cases when $\gamma = 0$ and when $\gamma = \pm 1$:
%\begin{align}
%    \sigma_\mathrm{PBH}^2 &= \frac{(q_\mathrm{max}-1)^2}{q_\mathrm{max}\log^2 q_\mathrm{max}} - 1, \\
%    \sigma_{\pm1}^2 &= \frac{\log q_\mathrm{max}}{2}\frac{q_\mathrm{max} + 1}{q_\mathrm{max} - 1} - 1.
%\end{align}

%\textcolor{red}{As expected, $\gamma = 0$ represents the widest possible bounded power law distribution for a fixed $q_\mathrm{max}$, scaling as $\sigma^2_0 \sim q_\mathrm{max}/\log^2q_\mathrm{max}$ for large maximum mass ratios. In contrast, $\gamma = \pm 1$ represent turning points where the width of the distribution has logarithmic or slower dependence on $q_\mathrm{max}$. Concretely this translates to $\sigma_0 \sim 4.46$ and $\sigma_{\pm1} \sim 1.57$ for $q_\mathrm{max} = 1000$. The widths for distributions with $|\gamma| < 1$ are bounded from above by $\sigma_0$ and from below by $\sigma_{\pm1}$, while $|\gamma| > 1$ distributions will have widths $\sigma < \sigma_{\pm1}$, approaching that of the monochromatic distribution at large $|\gamma|$ regardless of the value of $q_\mathrm{max}$.}

\section{Calculating the constraining observables}
\label{sec:observables}
\subsection{Expected detectable merger event rate}
\label{subsec:merger}
For resolvable PBH mergers, the expected number of merger events per year $N_\mathrm{det}$ is given by \cite{Raidal_2017, Dominik_2015} 
\begin{equation}
N_\mathrm{det} = \int\mathrm{d}z\mathrm{d}M_1\mathrm{d}M_2\,\frac{1}{1+z}\frac{\mathrm{d}V_c(z)}{\mathrm{d}z}\frac{\mathrm{d}^2R_\mathrm{PBH}}{\mathrm{d}M_1\mathrm{d}M_2}p_\mathrm{det}(M_1, M_2, z).
\label{eqref:ndet}
\end{equation}
Here, $V_c(z)$ is the comoving volume per unit redshift, and $p_\mathrm{det}$ is the binary detectability of any mass pair $M_1$ and $M_2$ at a redshift $z$ for some particular GW detector, and $R_\mathrm{PBH}$ is the differential PBH merger rate. $p_\mathrm{det}$ is determined by the ratio between the optimal detector signal-to-noise ratio (SNR) $\rho_\mathrm{det}$ and a preset detection threshold $\rho_\mathrm{thr}$, which we set to be equal to 8. We use the form of $p_\mathrm{det}$ as described in the Appendix of~\cite{Dominik_2015} for a single detector.
 
The differential PBH merger rate $R_\mathrm{PBH}$ is given by~\cite{Nakamura_1997, PhysRevD.96.123523, Raidal_2017}
\begin{align}
\frac{\mathrm{d}^2R_\mathrm{PBH}}{\mathrm{d}M_1\mathrm{d}M_2} &= \frac{1.6\times 10^6}{\mathrm{Gpc}^3\,\mathrm{yr}}f^{53/37}_\mathrm{PBH}\eta^{-34/37}\left(\frac{t}{t_0}\right)^{-34/37} \left(\frac{M_\mathrm{tot}}{M_\odot}\right)^{-32/37}\times \nonumber\\
&\times S(M_1, M_2, f_\mathrm{PBH}, t)\psi(M_1)\psi(M_2),
\label{eqref:merger rate}
\end{align}
where $M_1$ and $M_2$ are the principal masses of the PBH binary, $M_\mathrm{tot} = M_1 + M_2$ is the initial total mass of the binary upon formation, $\eta = (M_1M_2)/M_\mathrm{tot}^2$ is the symmetric mass ratio of the binary, $\psi(M)$ denotes the PBH mass distribution upon formation, and the suppression factor $S = S_1 \times S_2$, where $S_1$ accounts for disruption due to inhomogeneities in the DM fluid and neighboring PBHs and $S_2$ accounts for disruption due to substructures forming later in the Universe. We discuss each of these factors below.

An exact expression for the first suppression factor $S_1$ is given by Eq.~2.37 in Ref.~\cite{Raidal_2019} for extended mass functions. Ref.~\cite{Hutsi_2021} gives an approximation which we adopt here for efficient computation, given by
\begin{equation}
\label{eq:suppression}
    S_1 \approx 1.42\left[\frac{\sigma_\mathrm{PBH}^2 + 1}{\bar{N}(y) + C} + \frac{\sigma^2_M}{f^2_\mathrm{PBH}}\right]^{-21/74}e^{-\bar{N}(y)}, 
\end{equation}
where $\sigma^2_\mathrm{PBH}$ is the variance characterizing the spread of the PBH mass function (introduced in Sec.~\ref{subsubsec:dist_plaw}) and $\sigma^2_M$ is the rescaled variance of matter density perturbations at the time of formation of the binary, approximately equal to $3.6\times 10^{-5}$. Here, $y$ is the minimal distance to the next-nearest-neighbor (i.e., third PBH) and $\bar{N}(y)$ is the expected number of PBHs within a sphere of comoving radius $y$ around the initial PBH pair, estimated as
\begin{equation}
    \bar{N}(y) = \frac{M_\mathrm{tot}}{\langle M \rangle}\frac{f_\mathrm{PBH}}{f_\mathrm{PBH} + \sigma_M},
\end{equation}
and $C$ is a fitting parameter written out as
\begin{equation}
    C(f_\mathrm{PBH}, \sigma_\mathrm{PBH}) = f_\mathrm{PBH}^2\frac{\sigma^2_\mathrm{PBH} + 1}{\sigma_M^2}\left\{\left[\frac{\Gamma(29/37)}{\sqrt{\pi}}U\left(\frac{21}{74},\frac{1}{2},\frac{5f_\mathrm{PBH}^2}{6\sigma^2_M}\right)\right]^{-74/21} - 1\right\}^{-1},
\end{equation} 
where $U$ is the confluent hypergeometric function. We note that while this approximation is accurate to within $7\%$ for log-normal mass functions with widths $\sigma \leq 2$ according to Ref.~\cite{Hutsi_2021}, for the bounded power law distributions we study here, the deviations from the exact expression can generally be larger, albeit still acceptable for the parameter ranges of interest; we find that the deviation increases with increasing $\gamma$ and decreasing $M_{\rm tot}$. Specifically, for $\gamma = 10$ (the highest value we consider) and $M_{\rm tot} = 10^{-1} M_\odot$, the percent difference ranges from around 20--45\% for $f_\mathrm{PBH}$ from $10^{-4}$ to $10^{-3}$ and increases with $f_\mathrm{PBH}$ to around 50\% for $f_\mathrm{PBH} \gtrsim 10^{-2}$.

The second suppression factor $S_2$ accounts for PBH-binary scattering within DM structures (haloes/subhaloes) and an estimate was derived in Ref.~\cite{PhysRevD.101.043015} for a monochromatic mass distribution. For a given redshift of the merger, the value of $S_2$ depends only on $f_\mathrm{PBH}$, and is equal to 1 for $f_\mathrm{PBH}$ values up to some threshold value that depends on the redshift. For $z=0$, this threshold value attains its lowest value, around $0.003$; fit increases to around $0.03$ for $z=30$ and to around $0.1$ for $z=300$. In general, the impact of $S_2$ is not expected to be significant when $f_\mathrm{PBH}$ is low, as the probability of disruption is also lower. As far as we know, an expression valid for extended mass distributions is not yet available in the literature and investigating this could be the scope of future work. In this work, we have simply set $S_2 = 1$.

Note that the calculation of the expected event rate $N_\mathrm{det}$ involves both an integration across not just a redshift range but also across a mass range. For monochromatic PBH mass distributions, this mass range collapses to $M_c$; all detectable PBH merger events associated with the reference mass $M_c$ are due to binaries of total mass with the same order of magnitude as the reference mass $(M_\mathrm{tot} = 2M_c)$\footnote{We infer that the same can be said for narrowly-peaked lognormal mass distributions, as the probability of drawing a component mass much smaller or larger than $M_c$ falls off quite fast, reducing its weight towards the total integration.}. However, this is not the case for broad mass distributions such as power law. Instead, each event rate attributed to a power law distribution $\psi$ with parameters $M_c$ and $\gamma$ contains the integrated contribution of all binaries with component masses within the range of $M_\mathrm{min}$ to $M_\mathrm{max}$ sampled from $\psi$. We note that because of how the mass range bounds are defined in Sec.~\ref{subsec:mass_functions}, the resulting integration bounds in Eq.~\eqref{eqref:ndet} may technically include binaries with total masses outside the detection ranges we have set for LISA and ET in Sec.~\ref{subsec:waveforms}; in practice, we do not include these contributions for simplicity, as they are expected to be negligible. %\textcolor{blue}{This also allows us to potentially extend the mass window beyond what is projected by the monochromatic constraint, since $M_c$ values not in the monochromatic constraint window may still have a non-zero event rate being contributed to by other masses in the same distribution.}

\subsection{Expected stochastic GW background}
\label{sec:SGWB}
For a merging population of compact binary objects, with component masses $M_1$ and $M_2$, the generic expression for the expected SGWB spectrum at frequency $\nu$ is given by~\cite{phinney2001practicaltheoremgravitationalwave, Zhu_2011}
\begin{equation}
\label{eq:omega}
    \Omega_\mathrm{GW}(\nu) = \frac{\nu}{\rho_0} \int_0^{\frac{\nu_\mathrm{cutoff}}{\nu} - 1}\,\mathrm{d}z\mathrm{d}M_1\mathrm{d}M_2\,\frac{1}{(1+z)H(z)}\frac{\mathrm{d}^2R}{\mathrm{d}M_1\mathrm{d}M_2}\frac{\mathrm{d}E_\mathrm{GW}(\nu_S)}{\mathrm{d}\nu_S},
\end{equation}
where the present energy density $\rho_0 = 3H_0^2 c^2 / 8\pi G$, $H_0$ is the Hubble constant, $H(z)$ is the Hubble parameter at redshift $z$, $R$ is the differential merger rate, and $E_{\rm GW}$ is the energy spectrum of GWs from the binary black hole (BBH) mergers. Note that the redshift integration is done up to $z_\mathrm{cutoff} = (\nu_\mathrm{cutoff}/\nu) - 1$. We are interested in calculating Eq.~\ref{eq:omega} for PBH binary mergers, i.e., for $R = R_\mathrm{PBH}$ given in Eq.~\ref{eqref:merger rate}. We assume a $\Lambda$CDM cosmology with $h = 0.7$, $\Omega_m = 0.3$, and $\Omega_\mathrm{de} = 0.7$.

The energy spectrum of GWs from BBH mergers is given by a phenomenological formulation of $E_\mathrm{GW}$ in terms of the redshifted source frequency $\nu_S = \nu(1+z)$. We lay out here the full form of $E_\mathrm{GW}$ formulated in Ref.~\cite{PhysRevLett.106.241101}, utilizing the non-spinning ($\chi_1 = \chi_2 = 0$) limit:
\begin{equation}
\label{eq:e_gw}
    \frac{\mathrm{d}E_\mathrm{GW}(\nu)}{\mathrm{d}\nu} = \frac{(\pi G)^{2/3}M_\mathrm{tot}^{5/3}\eta}{3}
    \begin{cases}
        \nu^{-1/3}(1 + \alpha_2\Bar{\nu}^2)^2,&\quad\mathrm{if}\,\nu < \nu_\mathrm{merg}, \\
        w_1\nu^{2/3}(1 + \epsilon_1\Bar{\nu} + \epsilon_2\Bar{\nu}^2)^2,&\quad\mathrm{if}\,\nu_\mathrm{merg} \leq \nu < \nu_\mathrm{ring}, \\
        w_2\nu^{2}[\mathcal{L}(\nu, \nu_\mathrm{ring}, \varsigma)]^2,&\quad\mathrm{if}\,\nu_\mathrm{ring} \leq \nu < \nu_\mathrm{cutoff},
    \end{cases}
\end{equation}
where $\Bar{\nu} = (\pi GM_\mathrm{tot}\nu/c^3)^{1/3}$, $\alpha_2 = (451/168)\eta -323/224$, $\epsilon_1 = -1.8897$, and $\epsilon_2 = 1.6557$. The continuity constants $w_i$ connecting the different frequency regimes are given by
\begin{align}
    w_1 &= \nu_\mathrm{merg}^{-1}\frac{(1+\alpha_2\Bar{\nu}_\mathrm{merg}^{2})^2}{(1 + \epsilon_1\Bar{\nu}_\mathrm{merg}^{1} + \epsilon_2\Bar{\nu}_\mathrm{merg}^{2})^2} \nonumber\\
    w_2 &= w_1\nu_\mathrm{ring}^{-4/3}(1 + \epsilon_1\Bar{\nu}_\mathrm{ring}^{1} + \epsilon_2\Bar{\nu}_\mathrm{ring}^{2})^2,
\end{align}
where $\Bar{\nu}_i = (\pi GM_\mathrm{tot}\nu_i/c^3)^{1/3}$, and the frequencies $\nu_i = \{\nu_\mathrm{merg}, \nu_\mathrm{ring}, \nu_\mathrm{cutoff}, \varsigma\}$ are the merger, ringdown, and cutoff frequencies respectively of a BBH with parameters $M_\mathrm{tot}$ and $\eta$ for some specific waveform model, while $\varsigma$ is the width of the Lorentzian distribution $\mathcal{L}$. Writing out these frequencies, we obtain
\begin{align}
    \Bar{\nu}_\mathrm{merg} &=  (1 - 4.455 + 3.521) + 0.6437\eta - 0.05822\eta^2 - 7.092\eta^3, \nonumber\\
    \Bar{\nu}_\mathrm{ring} &=  (1 - 0.63)/2 + 0.1469\eta - 0.0249\eta^2 + 2.325\eta^3, \nonumber\\
    \pi GM_\mathrm{tot}\varsigma/c^3 &=  (1 - 0.63)/4 - 0.4098\eta + 1.829\eta^2 - 2.87\eta^3, \nonumber \\
    \Bar{\nu}_\mathrm{cutoff} &= 0.3236 - 0.1331\eta - 0.2714\eta^2 + 4.922\eta^3.
\end{align}

In order to derive constraints on $f_{\rm PBH}$, we need to set a detection threshold. Following~\cite{PhysRevD.88.124032}, we calculate the SNR for SGWB detection from
\begin{equation}
\label{eq:SGWB_SNR}
    \rho_\mathrm{det} = \sqrt{\mathcal{T}\int_{\nu_\mathrm{min}}^{\nu_\mathrm{max}}\mathrm{d}\nu\, \left(\frac{\Omega_\mathrm{GW}(\nu)}{\Omega_\mathrm{sens}(\nu)}\right)^2},
\end{equation}
where $\Omega_\mathrm{sens}(\nu)$ is the dimensionless energy in noise of the detector, $\mathcal{T} = 1\,\mathrm{yr}$ is the target observation period, and the limits of integration $[\nu_{\rm min}, \nu_{\rm max}]$ define the bandwidth of the detector. Here, we set our SNR threshold to be $\rho_\mathrm{thr} = 5$ and consider SGWB detection if $\rho_\mathrm{det} \geq \rho_\mathrm{thr}$.

Finally, we note that we have adopted the approximation in Eq.~\ref{eq:e_gw} and applied it to our calculations involving binaries with mass ratios up to $q_{\rm max}=1000$, even though Ref.~\cite{PhysRevLett.106.241101} notes that it is recommended only for modeling mergers up to mass ratios $q_\mathrm{max} \lesssim 10$, given limitations in the numerical relativity calculations performed by the authors. In Sec.~\ref{subsec:varying_qmax}, we present how our derived constraints on PBH abundance vary with the choice of $q_{\rm max}$.

%As we will show in Sec.~\ref{subsec:SGWB_constraint}, naively extending this model of $E_\mathrm{GW}$ up to $q_\mathrm{max} = 1000$ introduces no pathologies to the behaviour of $\Omega_\mathrm{GW}$. Given the limitation of the $E_\mathrm{GW}$ model, however, our projections may be understating the effect higher population $q_\mathrm{max}$ has on the constraint.

\section{Constraints on PBH abundance}
\label{sec:measurable}

\subsection{Methodology}
\label{subsec:method}

Before we present our results on constraints on PBH abundance from resolvable mergers and the SGWB (in Secs.~\ref{subsec:resolvable} and \ref{subsec:SGWB_constraint}, respectively), we first summarize the parameters assumed in our SNR calculations in Table~\ref{tab:detector_specs}.

\begin{table}[htbp!]
\centering
\begin{tabular}{|l|l|l|}
\hline
& Resolvable mergers & SGWB                                     \\\hline
Binary mass detection range $M_\mathrm{tot}$ & \begin{tabular}[c]{@{}l@{}}LISA: 1--$10^9\,M_\odot$\\ 
ET: $10^{-6}$--$10^6\,M_\odot$\end{tabular} & 
\begin{tabular}[c]{@{}l@{}}LISA:  $10^{-6}$--$10^9\,M_\odot$ \\ ET: $10^{-6}$--$10^6\,M_\odot$\end{tabular} \\\hline
Binary mass ratios $q$ & 1 -- 1000  & 1 -- 1000                                                                                                         \\\hline
Binary component spins $\chi_1$ and $\chi_2$ & 0                                                                                                 & 0                                                                                                                \\\hline
Waveform model                               & IMRPhenomXAS                                                                                      & Eq.~\ref{eq:e_gw}                                                                         \\\hline
Simulated observation time $\mathcal{T}$     & 1 yr                                                                                              & 1 yr                                                                                                             \\\hline
Detection threshold $\rho_\mathrm{thr}$      & 8                                                                                                 & 5    \\\hline                                                                                                           
\end{tabular}
\caption{Parameters assumed in the calculation of SNR for resolvable mergers and SGWB observations.}
%the setup of detectability $p_\mathrm{det}$ and SGWB detection.}
\label{tab:detector_specs}
\end{table}

With these assumptions, we obtain constraints for monochromatic and bounded power law mass distributions (as described in Sec.~\ref{subsubsec:dist_plaw}), with selected values of the power-law exponent $\gamma$. For resolvable mergers, we consider selected values of $\gamma$ within the range $[-10, 10]$, including $\gamma = 0$; specifically, we defined 7 equally log-spaced values of $|\gamma|$ (for a total of 15 values). For the SGWB, we consider 3 values of $\gamma$, equal to $-1/2,\,0,\,1/2$. We obtain constraints for a range of reference masses $M_c$ equal to the range in total masses listed in Table~\ref{tab:detector_specs}. We note that for simplicity, initial spin, clustering, and accretion effects on the PBH merger rate are ignored.

%mass distribution and merger rate calculations are as follows:
%\begin{itemize}
%    \item Mass functions: monochromatic a
%    \item Range in reference mass $M_c$: same as ranges in total mass $M_\mathrm{tot}$ (listed in Table~\ref{tab:detector_specs}).
%    \item Resolvable source power law exponent $\gamma$ range: $|\gamma| \in [0, 10]$ for both $\gamma < 0$ and $\gamma > 0$.
%    \item SGWB source $\gamma$ range: $\gamma = -1/2,\,0,\,1/2$ 
%    \item Power law cutoffs determined by Eq.~\ref{eq:M_c}.
%    \item Initial spin, clustering, and accretion effects on the PBH merger rate are ignored.
%\end{itemize}

For the calculation of the $f_\mathrm{PBH}$ constraints from resolvable mergers, we first define a log-spaced grid in $M_\mathrm{tot}$ and a linearly-spaced grid in $q$ and pre-calculate the SNR for each detector using \texttt{gwent}~\cite{Kaiser:2021}; the ranges in $q$ and $M_\mathrm{tot}$ (per detector) are as specified in Table~\ref{tab:detector_specs}. This results in a $50 \times 50$ $M_\mathrm{tot}-q$ SNR table per redshift, for 100 equal log-spaced bins from $z = 0$ to $300$. We do this for both LISA and ET, preparing a total of 100 SNR tables each.

We use these SNR values $\rho_\mathrm{det}$ to obtain $p_\mathrm{det}$ in Eq.~\eqref{eqref:ndet} for a given combination of $M_1$, $M_2$, and $z$. The bounds of mass integration are determined by $M_\mathrm{min}$ and $M_\mathrm{max}$, and we assign $q_\mathrm{max} = 1000$ unless specified. The bounds for redshift integration are set to be $z \in [0.01, 30]$ for low redshift events and $z \in [30, 300]$ for high redshift events. 

Next, we obtain the expected PBH integrated event rate $N_\mathrm{det}$ for each $(M_c, \gamma, f_\mathrm{PBH})$ point in the parameter space, where we choose 100 equal log-spaced bins in $M_c$ for the specified range (per detector) and $f_\mathrm{PBH}$ runs from $10^{-8}$ to $1$ in 100 equal log-spaced bins. 
%We consider 15 values of $\gamma$, 7 with positive and 7 with negative signs, alongside $\gamma = 0$. Note that $f_\mathrm{PBH}$ and $\gamma$ enter through the differential PBH merger rate given in Eq.~\eqref{eqref:merger rate}. 

Once all event rates within the parameter space have been calculated, we scan through the $f_\mathrm{PBH}$ range for each $(M_c, \gamma)$ pair and pick the minimum $f_\mathrm{{PBH}}$ value where the expected event rate $N_\mathrm{det} \geq 1$. These minimum $f_\mathrm{PBH}$ values serve as our constraint for a particular reference mass $M_c$ and power law distribution. 

We follow a similar procedure to calculate constraints from the SGWB spectrum, going through each point in the $(M_c, \gamma, f_\mathrm{PBH})$ parameter space and calculating the expected SGWB spectrum using Eq.~\ref{eq:omega}, then applying Eq.~\ref{eq:SGWB_SNR} to get the corresponding SNR, $\rho_\mathrm{det}$. Finally, we choose the minimum $f_\mathrm{PBH}$ value where this is higher than the threshold value, $\rho_\mathrm{det} \geq \rho_\mathrm{thr}$ (for a particular $M_c$ and $\gamma$).

\subsection{Constraints from resolvable mergers}
\label{subsec:resolvable}
\begin{figure}
    \centering
    \begin{subfigure}{\textwidth}
        \centering
        \includegraphics[width=0.9\textwidth]{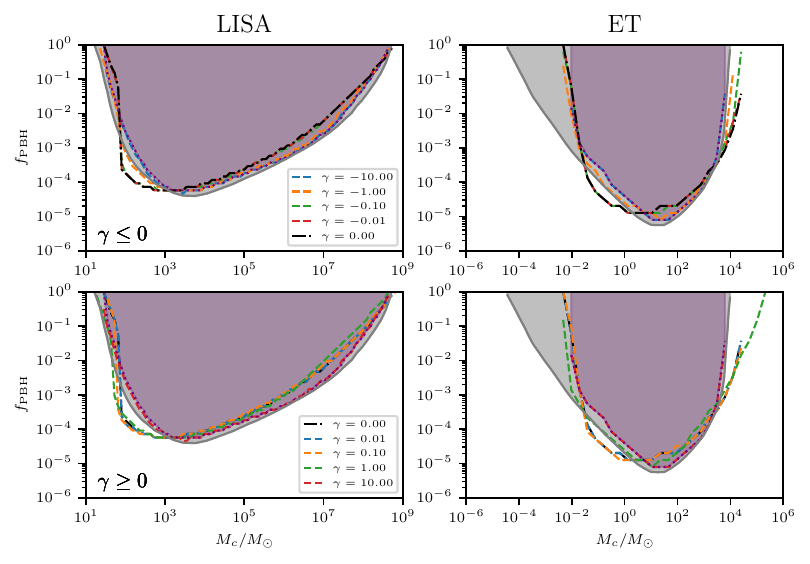}
        \caption{Low-redshift events ($z \in [0.01, 30]$)\label{fig:f-M_low}}
    \end{subfigure}
    
    \begin{subfigure}{\textwidth}
        \centering
        \includegraphics[width=0.9\textwidth]{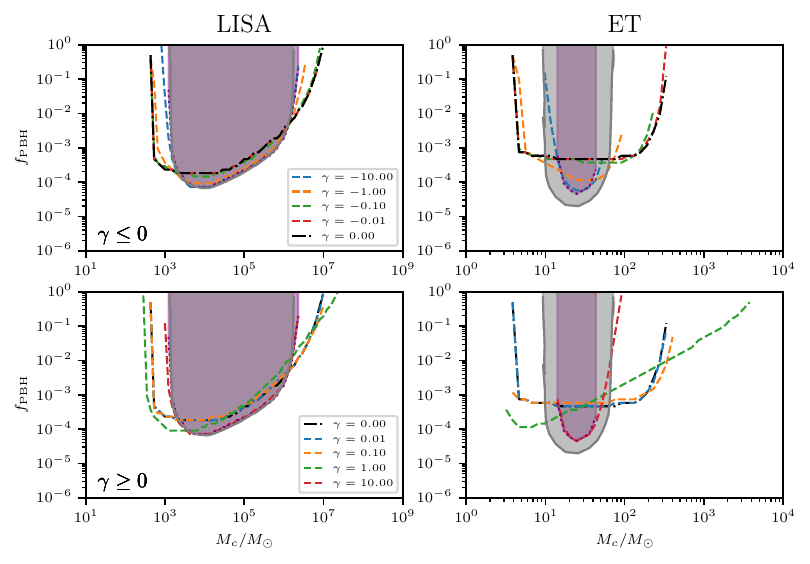}
        \caption{High-redshift events ($z \in [30, 300]$)\label{fig:f-M_high}}
    \end{subfigure}
    \caption{Minimum PBH abundance $f_\mathrm{PBH}$ constraint curves for bounded power law distributions with varying power-law exponents $\gamma$, for selected values $\gamma \leq 0$ (top panels) and $\gamma \geq 0 $ (bottom panels), for LISA (left panels) and ET (right panels). The purple regions show constraints derived for monochromatic distributions (using IMRPhenomXAS waveforms). For comparison, solid gray regions show the monochromatic constraints obtained by Ref.~\cite{De_Luca_2021} (using IMRPhenomD waveforms).\label{fig:f-M}}
\end{figure}

\begin{figure}
    \centering
    \begin{subfigure}{\textwidth}
        \centering
        \includegraphics[width=0.9\textwidth]{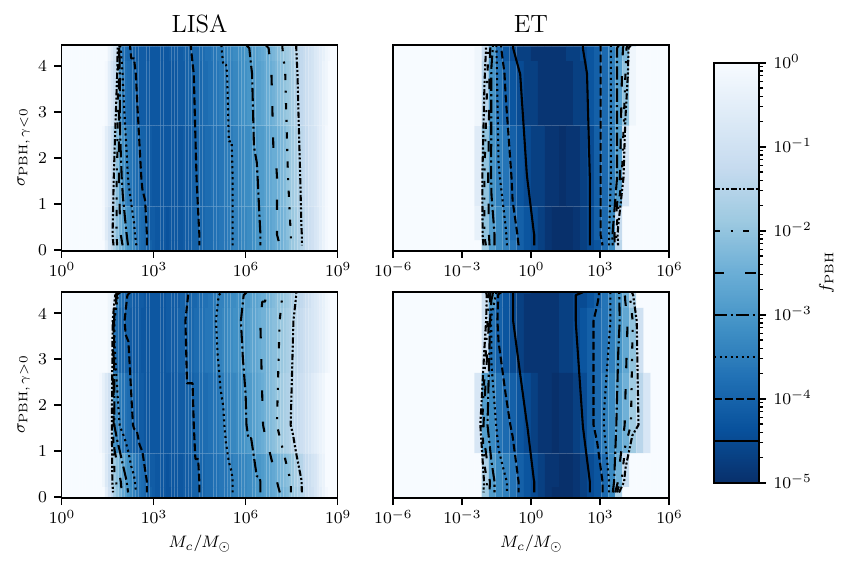}
        \caption{Low-redshift events ($z \in [0.01, 30]$)\label{fig:s-M_low}}
    \end{subfigure}
    
    \begin{subfigure}{\textwidth}
        \centering
        \includegraphics[width=0.9\textwidth]{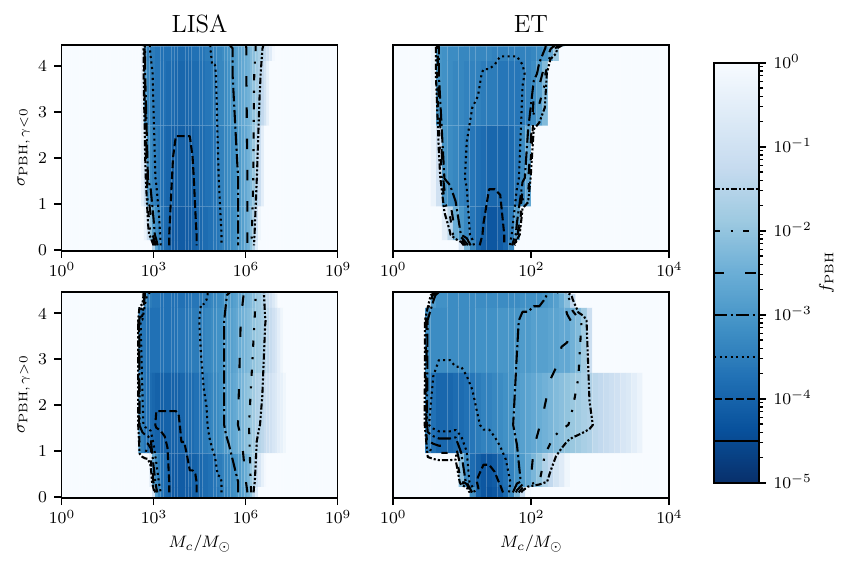}
        \caption{High-redshift events ($z \in [30, 300]$)\label{fig:s-M_high}}
    \end{subfigure}
    \caption{$M_c-\sigma_{\rm PBH}$ contour plots for LISA (left panels) and ET (right panels). Top panels show constraints for distributions with $\gamma \le 0$, while bottom panels show constraints for $\gamma \ge 0$. Values of the $f_\mathrm{PBH}$ contours are indicated in the legend on the color bars. Unshaded white regions are classified as unconstrained by the detector (corresponding to $f_{\rm PBH} \ge 1)$, while the darkest shades of blue signify regions of the parameter space with the lowest minimum PBH abundances.\label{fig:s-M}}
\end{figure}

We now present the constraints on the minimum PBH abundance $f_\mathrm{PBH}$ from GW detections of resolvable mergers. Figures~\ref{fig:f-M_low} \&~\ref{fig:f-M_high} show constraint curves as a function of reference mass $M_c$ for bounded power law distributions with varying power-law exponents $\gamma$, for low and high redshift events, respectively. In both Figures, the left and right panels show projected constraints for LISA and ET, respectively, and the top and bottom panels show constraints for distributions with $\gamma \leq 0$ and $\gamma \geq 0$, respectively. Constraints from monochromatic distributions are shown by the purple dotted lines and shaded regions for comparison. Finally, the solid gray regions show the monochromatic constraints obtained by Ref.~\cite{De_Luca_2021}. We find good agreement between the two monochromatic results for LISA, though not for ET, which may due to differences in assumptions, e.g., different choices in waveform model (IMRPhenomXAS in this work vs. IMRPhenomD in Ref.~\cite{De_Luca_2021}).

As another way to visualize how these constraints vary with varying distribution profiles, Figures~\ref{fig:s-M_low} \&~\ref{fig:s-M_high} show contour plots of $f_\mathrm{PBH}$ over the studied parameter space in $M_c$ and $\sigma_{\rm PBH}$. Recall that there is a one-to-one correspondence between the power-law exponent $\gamma$ and the width of the distribution $\sigma_{\rm PBH}$ and for our choice of $q_{\rm max}=1000$, the maximum value of $\sigma_{\rm PBH}$ is around 4.46, corresponding to $\gamma=0$, i.e., a top-hat distribution (c.f. Table~\ref{tab:sigma_gamma}). As in the previous set of Figures, the left and right panels show projected constraints for LISA and ET, respectively, and the top and bottom panels show constraints for distributions with $\gamma \leq 0$ and $\gamma \geq 0$, respectively. Dark blue regions represent regions of the parameter space where there is an appreciably lower minimum abundance constraint (i.e., $f_\mathrm{PBH} \lesssim 10^{-2}$), while the light blue regions correspond to regions where at least percent-level abundance is required to obtain a resolvable signal in one observation year. Regions colored white (unshaded) are classified as regions which require $f_\mathrm{PBH} \geq 1$, i.e., unconstrained by the detector; in other words, the detector cannot set a physical constraint for a mass distribution within this region of the parameter space.

For the low-redshift events (Figs.~\ref{fig:f-M_low} \&~\ref{fig:s-M_low}), the constraints for the minimum PBH abundance do not vary significantly from the monochromatic case for the extended mass distributions considered, for both LISA and ET. In contrast, for the high-redshift events (Figs.~\ref{fig:f-M_high} \&~\ref{fig:s-M_high}), we find that the mass window generally broadens with smaller $|\gamma|$ (higher $\sigma_{\rm PBH}$), corresponding to flatter (broader) distributions, for both LISA and ET. The minimum abundances generally become higher as well, up to one order of magnitude higher for ET at $|\gamma| \sim 0$ compared to the monochromatic case. Meanwhile, for larger $|\gamma|$ (lower $\sigma_{\rm PBH}$) values, the distributions approach a Dirac delta distribution, and the constraints approach the monochromatic case, as expected.

\begin{figure}
    \centering
    \includegraphics{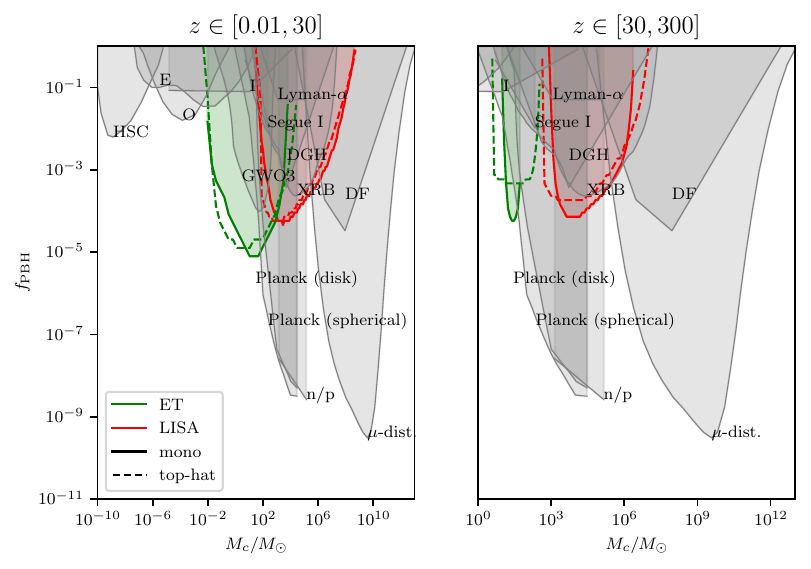}
    \caption{Projected $f_{\rm PBH}$ constraints for ET (green) and LISA (red) for monochromatic (solid) and top-hat (dashed) PBH mass distributions, from low- and high-redshift resolvable mergers (left and right panels, respectively), compared against existing present-day monochromatic constraints from the literature (solid curves, colored in grey). These include constraints from gravitational wave observations in the LVK O3 run (GWO3)~\cite{PhysRevD.110.023040} (shown in the left panel only), Planck CMB measurements~\cite{PhysRevResearch.2.023204}, dwarf galaxy dynamics (Segue I)~\cite{PhysRevLett.119.041102} and heating (DGH)~\cite{Lu_2021, Takhistov_2022}, Lyman-$\alpha$ observations~\cite{PhysRevLett.123.071102}, X-ray binary observations (XRB)~\cite{Inoue_2017}, dynamical friction (DF)~\cite{10.1093/mnras/sty1204}, the neutron-to-proton ratio (n/p)~\cite{PhysRevD.94.043527}, microlensing observations from Subaru HSC~\cite{Niikura2019,PhysRevD.101.063005}, MACHO/EROS (E)~\cite{Alcock_2001_1,Alcock_2001}, OGLE (O)~\cite{PhysRevD.99.083503}, and Icarus (I)~\cite{PhysRevD.97.023518}, as well as the CMB $\mu$-distortion~\cite{PhysRevD.97.043525}. Note that for the $\mu$-distortion constraint, we chose the FIRAS value corresponding to large non-Gaussianity $(p = 0.5)$.\label{fig:comparison}}
\end{figure}

Figure~\ref{fig:comparison} places our results in the context of existing constraints from the literature, which generally assume a monochromatic PBH mass distribution. Here, we show our derived constraints for ET (green) and LISA (red) for both monochromatic (solid) and top-hat (i.e., $\gamma=0$) (dashed) distributions. This value of $\gamma$ corresponds to the most contrasting departure from the monochromatic case. We compare these against constraints from gravitational-wave observations in the LIGO-Virgo-KAGRA (LVK) O3 run (GWO3)~\cite{PhysRevD.110.023040}, Planck CMB measurements (assuming spherical and disk accretion)~\cite{PhysRevResearch.2.023204}, dwarf galaxy dynamics (Segue I)~\cite{PhysRevLett.119.041102} and heating (DGH)~\cite{Lu_2021, Takhistov_2022}, Lyman-$\alpha$ observations~\cite{PhysRevLett.123.071102}, X-ray binary observations (XRB)~\cite{Inoue_2017}, dynamical friction (DF)~\cite{10.1093/mnras/sty1204}, the neutron-to-proton ratio (n/p)~\cite{PhysRevD.94.043527}, microlensing observations from Subaru HSC~\cite{Niikura2019,PhysRevD.101.063005}, MACHO/EROS (E)~\cite{Alcock_2001_1,Alcock_2001}, OGLE (O)~\cite{PhysRevD.99.083503}, and Icarus (I)~\cite{PhysRevD.97.023518}, and CMB $\mu$-distortion~\cite{PhysRevD.97.043525}, shown as the grey regions on the plots. We show the same regions on both low and high redshift plots for the sake of easy comparison, although we acknowledge that some of these constraints may be relaxed at high redshift. We note that each of these existing constraints have varying degrees of uncertainty and assumptions. In addition, these constraints are expected to change depending on the shape of the mass distribution considered~\cite{PhysRevD.96.023514}.

We highlight that, as previously reported by Ref.~\cite{De_Luca_2021}, future GW observatories open a window to detect high-redshift PBHs, as illustrated in the right panel of Figure~\ref{fig:comparison}. These constraints exist even if the population at $z < 30$, which can be confused with an astrophysical BH population, is neglected, providing possible smoking-gun evidence for PBHs. LISA and ET observations would complement each other by probing different mass windows and this remains true for the case of extended mass distributions studied here. %extended regardless of power law shape. %albeit the constraints become less stringent for increasingly positive power law exponents. 

At low redshift, ET projections generally lower the upper bound of abundance set by LVK GWO3. On the other hand, low-redshift LISA projections overlap with existing constraints from local observations, particularly those coming from present-day constraints on PBH accretion (DGH, XRB)~\cite{Lu_2021, Takhistov_2022, Inoue_2017} and dynamical effects on dwarf galaxies (Segue I)~\cite{PhysRevLett.119.041102}. These local constraints may potentially be relaxed at high redshift, or notably even disappear in the case of the LVK GWO3 constraints. However, going to high redshift does not let us evade constraints set by CMB $\mu$-distortion~\cite{PhysRevResearch.2.023204, PhysRevD.97.043525} and BBN observations (``n/p'')~\cite{PhysRevD.94.043527}, which most strongly constrain the region with $M_c$ above $10^2\,M_\odot$. We remark that even with a choice of a particularly conservative form of the $\mu$-distortion constraints (with non-Gaussianity parameter $p=0.5$), there is little to no room for PBHs of mass $\sim 10^6\,M_\odot$ at abundances $f_\mathrm{PBH} \sim 10^{-4}$. 

We note that Ref.~\cite{PhysRevD.96.023514} reports how the monochromatic constraints from the Segue I and Planck CMB shift for extended mass distributions, although these results are not shown here to avoid overcrowding the figures. Our contribution to the literature is to show how the monochromatic constraints from GW observations (as reported in e.g., Ref.~\cite{De_Luca_2021}) are expected to shift for extended mass distributions.

\subsection{Constraints from SGWB}
\label{subsec:SGWB_constraint}
\begin{figure}
    \centering
    \includegraphics[width=0.9\linewidth]{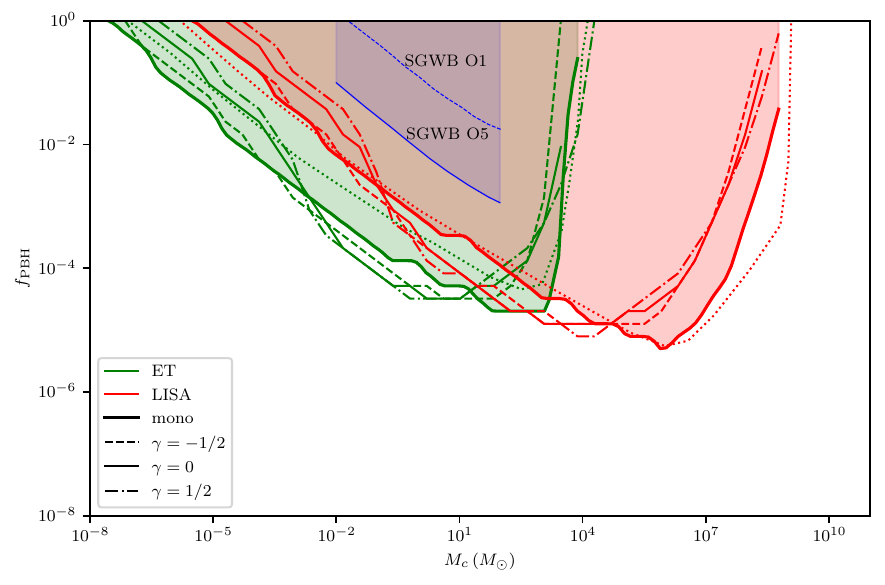}
    \caption{Minimum PBH abundance $f_\mathrm{PBH}$ constraint curves from SGWB for ET (green) and LISA (red) for monochromatic (thick solid) and top-hat (dashed) PBH mass distributions, compared against existing monochromatic constraints reported in Ref.~\cite{De_Luca_2021} (dotted). Also shown are the existing LIGO O1 SGWB constraint (blue dotted curve, labeled SGWB O1) and the projected O5 SGWB constraint~\cite{PhysRevLett.120.191102} (blue solid curve and region, labeled SGWB O5).} %update to differentiate between curves
    \label{fig:SGWB_mono_gamma}
\end{figure}

Next, we present the projected constraints on the minimum PBH abundance $f_{\rm PBH}$ from the unresolved SGWB. Figure~\ref{fig:SGWB_mono_gamma} shows the constraint curves for ET and LISA (green and red curves, respectively), for the monochromatic case (thick solid) together with the top-hat (i.e., $\gamma=0$) (thin solid) and bounded power-law distributions with $\gamma = -1/2$ (dashed) and $1/2$ (dot-dashed). We find that the constraints do not vary significantly for the extended distributions, except for a slight shift to the left compared to the monochromatic case, for both ET and LISA. Also shown for comparison are existing monochromatic constraints reported by Ref.~\cite{De_Luca_2021} (dotted curves), which seem to match qualitatively although they may not match exactly because of a difference in assumptions, including in the selected SNR detection thresholds.

The planned detectors are expected to broaden the mass windows probed and lower the abundance constraints by around an order of magnitude for LISA and 2 orders for ET compared to the projected O5 SGWB constraint, which itself is an improvement of around an order of magnitude compared with the existing LIGO O1 SGWB constraint (both reported in Ref.~\cite{PhysRevLett.120.191102}).

%Figure~\ref{fig:SGWB_mono_gamma} shows how our monochromatic constraint projections from the SGWB change with varying $\gamma$ for power law distributions. As with the cases at low redshift in Sec.~\ref{subsec:resolvable}, there are parts of the constraint shifted lower relative to the respective monochromatic constraint, such as the $1-10^4\,M_\odot$ mass window for LISA and the $10^{-3}-10\,M_\odot$ mass window for ET. For other parts of the constraint window, the required abundance is elevated. Generally speaking, given our current models, we can say that the projected SGWB constraints are only minimally responsive to changes in the mass distribution considered.

%\textcolor{blue}{Comparing our results with the ET and LISA SGWB constraints from Ref.~\cite{De_Luca_2021}, our monochromatic ET constraints are slightly more stringent. Compared to current and projected SGWB constraints from LIGO~\cite{PhysRevLett.120.191102}, however, ET and LISA SGWB observations still have the potential to significantly constrain the subsolar mass regime for PBHs.}

%Mention Figure from ref. for comparison; shape is similar, different magnitude, which could be due to different choice in SNR threshold

\subsection{Effect of varying maximum mass ratios}
\label{subsec:varying_qmax}

As noted in Sec.~\ref{subsec:waveforms}, in this work, we set the maximum mass ratio $q_\mathrm{max}=1000$ as constrained by the limits of the validity of our assumed waveform approximant, IMRPhenomXAS. As a test of systematics, we have checked how our derived constraints depend on the choice of maximum mass ratio $q_\mathrm{max}$. Varying this parameter modifies the value of the normalization of the mass function $\mathcal{N}_\mathrm{PL}$ (in Eq.~\ref{eqref:pl}), narrows the bounds of integration across the mass terms in the expected number of merger events $N_\mathrm{det}$ (in Eq.~\ref{eqref:merger rate}), and changes the value of the width parameter $\sigma_\mathrm{PBH}$ (consequently, affecting suppression factor $S_1$, given in Eq.~\ref{eq:suppression}).

\begin{figure}
    \centering
    \includegraphics[width=0.9\textwidth]{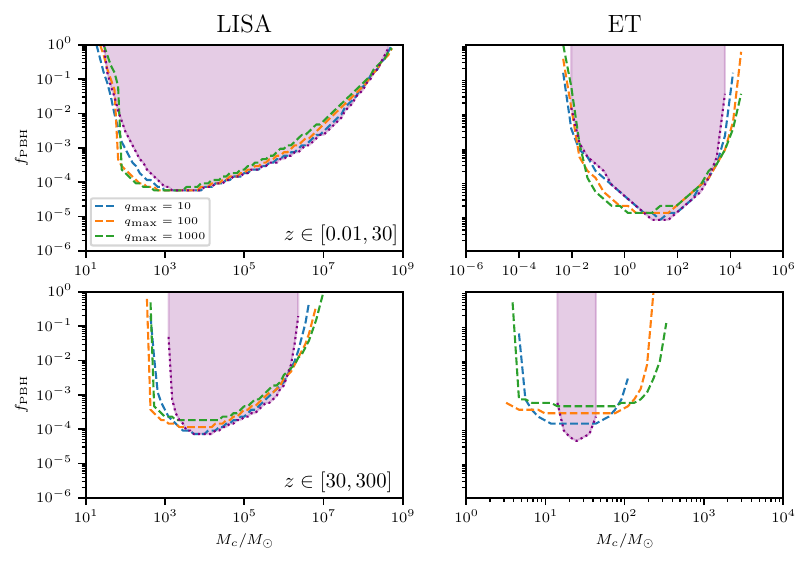}
    \caption{Minimum PBH abundance $f_\mathrm{PBH}$ constraint curves for LISA and ET (left and right panels) from low- and high-redshift events (top and bottom panels) assuming a top-hat distribution ($\gamma = 0$) for different choices of the maximum mass ratio $q_\mathrm{max} = 10, 100, 1000$ (blue, orange, and green dashed curves, respectively). For comparison, the monochromatic cases are shown by the dotted purple curves and regions.\label{fig:f-M_qmax}}

    \includegraphics[width=0.9\linewidth]{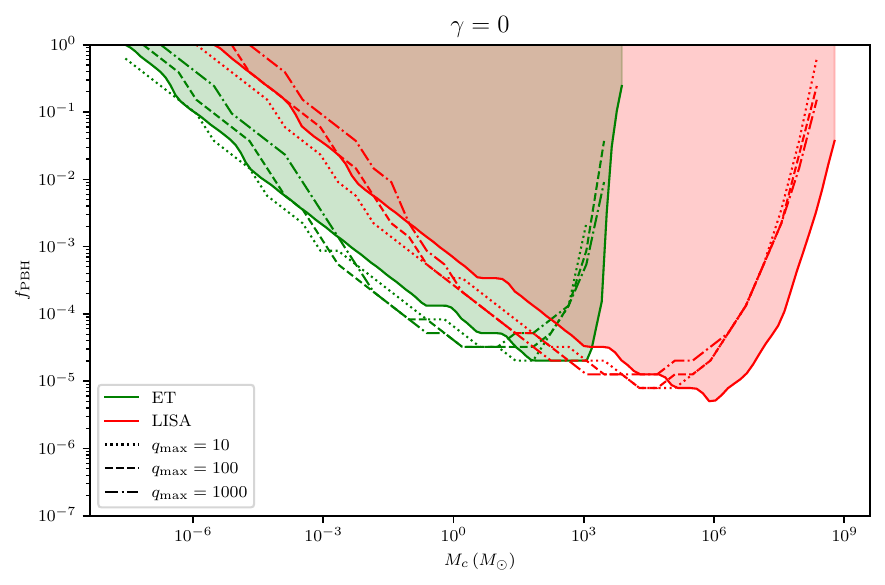}
    \caption{Minimum PBH abundance $f_\mathrm{PBH}$ constraint curves from SGWB for LISA (red) and ET (green) assuming a top-hat distribution ($\gamma = 0$) for different choices of the maximum mass ratio $q_\mathrm{max} = 10, 100, 1000$ (dotted, dashed, and dot-dashed curves, respectively). For comparison, the monochromatic cases are shown by solid curves and regions.}
    \label{fig:SGWB_mono_qmax}
\end{figure}

We do these checks for the case of the top-hat PBH mass distribution ($\gamma = 0$), which corresponds to the broadest profile studied and the largest value of $\sigma_\mathrm{PBH}$. Figures~\ref{fig:f-M_qmax} and~\ref{fig:SGWB_mono_qmax} show the results for constraints from resolvable mergers and SGWB, respectively. These show that the broadening of the mass windows observed in the high-redshift merger events is robust to the choice of $q_\mathrm{max}$ (lower panels of Fig.~\ref{fig:f-M_qmax}). Similarly, the slight shift in the left of the SWGB constraint curves is also consistently observed for lower $q_\mathrm{max}$ values. Overall, these checks show that the main findings we have presented in the previous subsections are robust to the particular choice of $q_\mathrm{max}$.

%\begin{figure}
%    \centering
%    \includegraphics[width=0.8\linewidth]{SNR-q_figures/SNR-qmax-varying_mass.pdf}
%    \caption{Detector SNRs as a function of total mass $M_\mathrm{tot}$ integrated up to fixed $q_\mathrm{max} = 1,\,10,\,100,\,1000$, for low (top panels) and high (bottom panels) redshifts. Left panels show ET SNRs, while right panels show SNRs for LISA. The shaded region represents $\mathrm{SNR} \leq 8$, lower than the threshold for a significant detection.}
%    \label{fig:SNR_qmax}
%\end{figure}

%\textcolor{blue}{For resolvable mergers, the response to changing $q_\mathrm{max}$ varies between redshift and detector design. The difference between constraints at low redshift is minimal, although the constraint with $q_\mathrm{max} = 10$ much more closely follows the monochromatic constraint compared to ones with higher mass ratios. We start seeing slightly more pronounced differences at higher redshift, although the ET constraints show a larger difference compared to the ones from LISA. In particular, the $q_\mathrm{max} = 10$ ET constraint remains fairly wide relative to the monochromatic case despite the distributions that produce it having only a width of $\sigma_\mathrm{PBH} \approx 0.73$. Finally, we note that changing $q_\mathrm{max}$ does not significantly affect the SGWB constraint, at best only changing the shape of the low mass end before the curve reaches its absolute minimum.}

\section{Conclusions and outlook}
\label{sec:conclu}

In this paper, we set projected $f_\mathrm{PBH}$ abundance constraints from simulated LISA and ET observations for extended PBH mass distributions from both resolvable mergers and the SGWB. We achieved this by considering broad power law distributions for a range of negative and positive exponents $\gamma$, as well as top-hat distributions $(\gamma = 0)$, considering binary mergers with mass ratios up to $q_\mathrm{max} = 1000$, and modeling these resolvable mergers using IMRPhenomXAS waveforms. We find that the most significant impact of extended mass distributions is observed for the high-redshift merger events, for which the minimum abundances are elevated (i.e., constraints are looser) and the probed mass windows broaden compared to the case where the distribution is assumed to be monochromatic. For the low-redshift merger events and SGWB, the constraints are not significantly affected.

Our results also demonstrate that the ability of the future GW observatories to detect high-redshift GW mergers, where astrophysical BHs are not expected to contribute, is robust to the simplifying assumption of a monochromatic PBH mass distribution. LISA and ET observations would complement each other by probing different PBH mass windows and this remains true for the case of the extended mass distributions we investigated.

In this work, we have not yet properly included the impact of considering the suppression of the constraints from low-redshift merger events arising from an astrophysical foreground, and this can be the subject of future work. Bayesian analysis accounting for the combined PBH and astrophysical BH populations have been performed for ground-based detectors, e.g., in Refs.~\cite{PhysRevD.102.123524}, \cite{Luca_2020}, and~\cite{PhysRevD.110.023040}, and extending this to future detectors would provide new and interesting insights.

Keeping assumptions about our PBH models constant, one way to evade the existing constraints is to have a detector that could probe at either higher sensitivities (reducing $f_\mathrm{PBH}$ necessary for a detection) or at higher mass ranges ($M \gtrsim 10^{11}\,M_\odot$) across a wide range of redshifts. Setting aside the feasibility of such a detector, however, evading the $\mu$-distortion bound by simultaneously lowering the required abundance and raising the PBH mass would quickly run us into the incredulity limit, which sets the abundance lower bound for $M \sim 10^{11}\,M_\odot$ at $f_\mathrm{PBH} \sim 10^{-10}$~\cite{10.1093/mnras/sty1204}. 

There are other, less detector-dependent approaches to evading the existing constraints. Modifying the model to account for phenomenological uncertainties such as clustering~\cite{De_Luca_2021, Young_2020, Ballesteros_2018} and accretion~\cite{PhysRevD.100.083016,PhysRevD.102.043505} may also provide an avenue for evasion. Incorporating broadly extended mass distributions to these model additions may however introduce new uncertainties, such as the implications of an extended distribution to the initial clustering, as well as potential changes to the local PBH accretion environments as a result of strong initial clustering. Recent developments in excursion set methods to model PBH initial clustering~\cite{PhysRevD.109.123538} may allow for a more nuanced phenomenological description of the average neighbor count $\bar{N}(y)$, accounting not only for total mass $M_\mathrm{tot}$ and average mass $\langle M \rangle$ but also the allowed mass ratios $q$ of the distribution. Applying this formalism to concretely describe initial clustering for power law mass functions is non-trivial. However, if accomplished, it may lead to the development of PBH merger rate suppression factors that apply to a more general range of PBH mass functions.

%The first is to consider subsolar ($M < 1\,M_\odot$) mergers, as well as potential constraints derived from the stochastic GW background. Both are beyond the scope of this current study, although they are worth investigating in the context of extended PBH mass distributions.

Finally, we remark that broadly extended mass distributions may be relevant when considering the PBH merger rate of initially three-body configurations~\cite{PhysRevD.101.043015, raidal2024formation}. This merger rate is considered subdominant to the the two-body merger rate in the regime considered by LVK~\cite{PhysRevD.110.023040}, but their analysis is restricted to narrow $(\sigma \lesssim 1.2)$ mass distributions. For these investigations, it would be interesting to consider even flatter and broader distributions. Generally, more work is needed in order to establish the nuance of how these very broad mass distributions affect the overall PBH merger rate.

\acknowledgments

We would like to thank Gabriele Franciolini, Andrew Kaiser, and Reginald Bernardo for responding to our questions regarding parts of this manuscript, as well as the anonymous referee for their insightful comments. We acknowledge the use of Python package \texttt{gwent}~\cite{Kaiser:2021} in the SNR and sensitivity curve computation for both LISA and ET.
%This is the most common positions for acknowledgments. A macro is
%available to maintain the same layout and spelling of the heading.

%\paragraph{Note added.} This is also a good position for notes added
%after the paper has been written.

% Bibliography

%% [A] Recommended: using JHEP.bst file
%% \bibliographystyle{JHEP}
%% \bibliography{biblio.bib}

%% or
%% [B] Manual formatting (see below)
%% (i) We suggest to always provide author, title and journal data or doi:
%% in short all the informations that clearly identify a document.
%% (ii) please avoid comments such as "For a review'', "For some examples",
%% "and references therein" or move them in the text. In general, please leave only references in the bibliography and move all
%% accessory text in footnotes.
%% (iii) Also, please have only one work for each \bibitem.

\bibliographystyle{JHEP}
\bibliography{main}

\end{document}